\documentclass{article}
\usepackage{titlesec}
\usepackage{graphicx}
\usepackage{arxiv}
\usepackage[utf8]{inputenc}
\usepackage{setspace}
\usepackage{geometry}
\usepackage{indentfirst}
\usepackage{amsmath}
\usepackage{booktabs,tabularx,makecell}
\usepackage{multirow}
\newcolumntype{Y}{>{\raggedright\arraybackslash}X}
\newcolumntype{L}{>{\raggedright\arraybackslash}X}
\usepackage{siunitx}
\usepackage[numbers]{natbib}
\usepackage[T1]{fontenc}
\usepackage{hyperref}
\usepackage{url}
\usepackage{amsfonts}
\usepackage{nicefrac}
\usepackage{microtype}
\usepackage{listings}
\usepackage{doi}

\title{HEAS: Hierarchical Evolutionary Agent Simulation Framework for Cross-Scale Modeling and Multi-Objective Search}
\date{}

\author{
\href{https://orcid.org/0000-0002-0883-4574}{\includegraphics[scale=0.06]{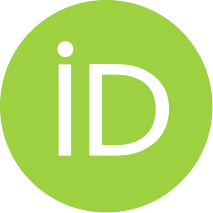}\hspace{1mm}Ruiyu Zhang}\\
Department of Politics and Public Administration\\
The University of Hong Kong\\
ruiyuzh@connect.hku.hk\\
\And
\href{https://orcid.org/0000-0002-0275-117X}{\includegraphics[scale=0.06]{orcid.pdf}\hspace{1mm}Lin Nie}\\
Department of Applied Social Sciences\\
The Hong Kong Polytechnic University\\
lin-apss.nie@polyu.edu.hk\\
\And
\href{https://orcid.org/0009-0005-7399-109X}{\includegraphics[scale=0.06]{orcid.pdf}\hspace{1mm}Xin Zhao}\\
Department of Applied Social Sciences\\
The Hong Kong Polytechnic University\\
xinnn.zhao@connect.polyu.edu.hk
}

\renewcommand{\headeright}{ }
\renewcommand{\undertitle}{ }
\renewcommand{\shorttitle}{Hierarchical Evolutionary Agent Simulation (HEAS)}

\hypersetup{
    pdftitle={HEAS: Hierarchical Evolutionary Agent Simulation Framework for Cross-Scale Modeling and Multi-Objective Search},
    pdfauthor={Ruiyu Zhang, Lin Nie, Xin Zhao},
    pdfkeywords={Agent-based Modeling, Scientific Computing, Evolutionary Optimization, Simulation Modeling, Multi-Objective Search},
}

\begin{document}
\maketitle

\begin{abstract}
HEAS is a Python framework that connects agent-based simulation, evolutionary search, and scenario-based evaluation in a single reproducible pipeline. It is designed for researchers who study systems where local interactions produce system-level outcomes---ecosystems, organizations, markets, or regulatory environments---and who need to search over candidate strategies and compare them across uncertain scenarios. HEAS combines three modules: a hierarchy runtime for composing simulations from reusable process layers, an evolutionary tuner for single- or multi-objective search backed by DEAP, and a game module for evaluating strategies across scenario ensembles. Its central design principle is the \emph{metric contract}: the same outcome function is shared by optimization, evaluation, and validation, so that different parts of an analysis cannot silently rank strategies by different quantities.
\end{abstract}


\section{Summary}
HEAS is a Python framework that connects agent-based simulation, evolutionary search, and scenario-based evaluation in a single reproducible pipeline. It is designed for researchers who study systems where local interactions produce system-level outcomes---ecosystems, organizations, markets, or regulatory environments---and who need to search over candidate strategies and compare them across uncertain scenarios. HEAS combines three modules: a hierarchy runtime for composing simulations from reusable process layers, an evolutionary tuner for single- or multi-objective search backed by DEAP, and a game module for evaluating strategies across scenario ensembles. Its central design principle is the \emph{metric contract}: the same outcome function is shared by optimization, evaluation, and validation, so that different parts of an analysis cannot silently rank strategies by different quantities.

\section{Statement of Need}
Many research problems now require simulation plus search. Ecologists compare management rules under uncertain shocks; social scientists test institutional designs in heterogeneous populations; engineers tune systems with competing performance and robustness objectives. Agent-based and individual-based models are now used across disciplines to represent complex systems made up of autonomous entities \citep{grimm2006odd}. These studies often need the same sequence of steps: build a model, run a search algorithm, evaluate candidate solutions across scenarios, and validate the final ranking.

Several mature tools cover parts of this workflow. Mesa \citep{mesa3_2025} and NetLogo \citep{wilensky1999} provide strong foundations for building and exploring agent-based models, but leave multi-objective search and scenario comparison to project-specific scripts. AgentPy \citep{agentpy2021} adds parameter sweeps and Monte Carlo experiments in Python, but does not connect these to evolutionary optimization or tournament evaluation. DEAP \citep{deap2012} offers a flexible evolutionary computation toolkit, but requires the user to wire up simulation coupling, metric definitions, and scenario handling manually. EMA Workbench \citep{kwakkel2017} focuses on exploratory modeling and robust decision making under deep uncertainty, emphasizing many-model experiments rather than multi-objective search over a single simulation hierarchy. OpenMOLE \citep{reuillon2013openmole} provides workflow-based model exploration with high-performance computing support, targeting large-scale parameter space exploration rather than integrated simulation-optimization with shared metric definitions.

In each case, the connection between simulation, search, and evaluation is left to the user. A common pattern is to couple an agent-based model to NSGA-II \citep{deb2002}, then re-score the resulting strategies in held-out scenarios. The optimizer, tournament evaluator, and validation code must compute the same outcome, yet this consistency is rarely enforced by the framework. Small differences in aggregation can change which strategy appears best without raising an error. HEAS is designed for researchers who need this end-to-end pipeline to be explicit, reproducible, and reusable across substantive domains.

\section{State of the Field}
Several frameworks address overlapping parts of the simulation-optimization pipeline. Table~\ref{tab:comparison} summarizes how HEAS positions itself relative to existing tools.

\begin{table}[htbp]
\centering
\caption{Capability comparison across simulation-optimization frameworks. ${}^a$OpenMOLE orchestrates external ABM models (e.g.\ from NetLogo) but does not provide agent classes or scheduling primitives. ${}^b$EMA Workbench supports exploratory scenario analysis and scenario discovery, but uses many-model ensembles rather than voting-rule-based tournaments. ``Shared metric across pipeline'' means the same outcome function is enforced by the framework across optimization, evaluation, and validation. ``Hierarchical composition'' means simulations can be assembled from layered, reusable process components.}
\label{tab:comparison}
\begin{tabular}{lccccccc}
\toprule
\textbf{Capability} & \textbf{Mesa} & \textbf{NetLogo} & \textbf{AgentPy} & \textbf{DEAP} & \textbf{EMA WB} & \textbf{OpenMOLE} & \textbf{HEAS} \\
\midrule
Agent-based modeling    & \checkmark & \checkmark & \checkmark & --- & --- & ---${}^a$ & \checkmark \\
Multi-objective search  & --- & --- & --- & \checkmark & --- & --- & \checkmark \\
Scenario analysis       & --- & --- & --- & --- & \checkmark${}^b$ & --- & \checkmark \\
Shared metric           & --- & --- & --- & --- & --- & --- & \checkmark \\
Hierarchical composition& --- & --- & --- & --- & --- & --- & \checkmark \\
Parallel / HPC          & --- & --- & --- & --- & --- & \checkmark & partial \\
Python-native           & \checkmark & --- & \checkmark & \checkmark & \checkmark & --- & \checkmark \\
\bottomrule
\end{tabular}
\end{table}

Mesa \citep{mesa3_2025} and NetLogo \citep{wilensky1999} are strong choices for building and exploring agent-based models, but they do not provide integrated multi-objective search or scenario-based evaluation. AgentPy \citep{agentpy2021} adds parameter sweeps and Monte Carlo experiments, but does not connect these to evolutionary optimization. DEAP \citep{deap2012} provides modular evolutionary algorithms; HEAS builds on DEAP rather than replacing it. EMA Workbench \citep{kwakkel2017} supports exploratory modeling and robust decision making under deep uncertainty, but frames the workflow around many-model ensembles rather than multi-objective search within a single simulation hierarchy. OpenMOLE \citep{reuillon2013openmole} orchestrates model exploration across high-performance computing resources, but does not enforce metric consistency across the optimization and evaluation stages.

HEAS contributes the missing integration layer: it connects hierarchical simulation, multi-objective search, scenario comparison, and a shared outcome definition in one framework. This makes HEAS useful to modelers who already know how to write a simulation, but need a more reliable bridge from simulation to search and comparative evaluation.

\section{Software Design}
HEAS has three modules that share a common data flow.

\textbf{Hierarchy runtime.} A \emph{stream} is a user-defined process that reads from and writes to a shared simulation context. Streams are the basic units of composition: each one encapsulates a piece of domain logic (an ecological process, a firm, a regulator, or an ODE system) while remaining agnostic about how other streams behave. A \emph{layer} groups streams that execute at the same temporal resolution. A \emph{hierarchy} orders layers so that slower processes (e.g.\ seasonal dynamics) frame faster ones (e.g.\ daily agent decisions). The following example defines a stream and registers it in a two-layer hierarchy:

\begin{lstlisting}[language=Python]
from heas.hierarchy.base import Stream, Context
from heas.hierarchy.orchestrator import (
    LayerSpec, StreamSpec, make_model_from_spec,
)

class Prey(Stream):
    def __init__(self, name, ctx, growth=0.05, **kw):
        super().__init__(name, ctx)
        self.growth = growth
        self.pop = 100.0

    def step(self):
        self.pop *= 1 + self.growth
        self.ctx.data["prey"] = self.pop

class Predator(Stream):
    def __init__(self, name, ctx, conversion=0.02, **kw):
        super().__init__(name, ctx)
        self.conversion = conversion

    def step(self):
        prey = self.ctx.data.get("prey", 0)
        self.ctx.data["predators"] = prey * self.conversion

spec = [
    LayerSpec(streams=[StreamSpec("prey", Prey, {"growth": 0.05})]),
    LayerSpec(streams=[StreamSpec("pred", Predator, {"conversion": 0.02})]),
]
model_factory = make_model_from_spec(spec, seed=42)
\end{lstlisting}

The hierarchy runtime executes layers in order, passing the shared context downstream. This separation means the same Prey stream can be reused in different hierarchies without modification.

\textbf{Evolutionary tuner.} The tuner connects DEAP-backed single- and multi-objective search (NSGA-II \citep{deb2002}, MOEA/D \citep{zhang2007moead}) to the hierarchy. A \emph{gene schema} maps candidate parameter vectors to stream parameters, and a single \texttt{metrics\_episode()} callable evaluates each episode. Deterministic seeding and parallel episode execution ensure reproducibility.

\textbf{Game module.} The game module evaluates candidate strategies across scenario ensembles and aggregates results using configurable voting rules (argmax, majority, Borda count, Copeland pairwise majority). It receives the same \texttt{metrics\_episode()} used by the tuner, so rankings are guaranteed to be computed from the same outcome definition.

The main design trade-off is that HEAS keeps domain behavior in user-defined streams rather than imposing a single scientific model type. This is less prescriptive than a domain-specific simulator, but it lets the same workflow serve ecology, organizational analysis, policy design, and mathematical systems modeling. The metric contract is the corresponding constraint: users define the outcome once, and the same definition is passed through search, tournament evaluation, and validation.

\begin{figure}[htbp]
\centering
\includegraphics[width=0.82\linewidth]{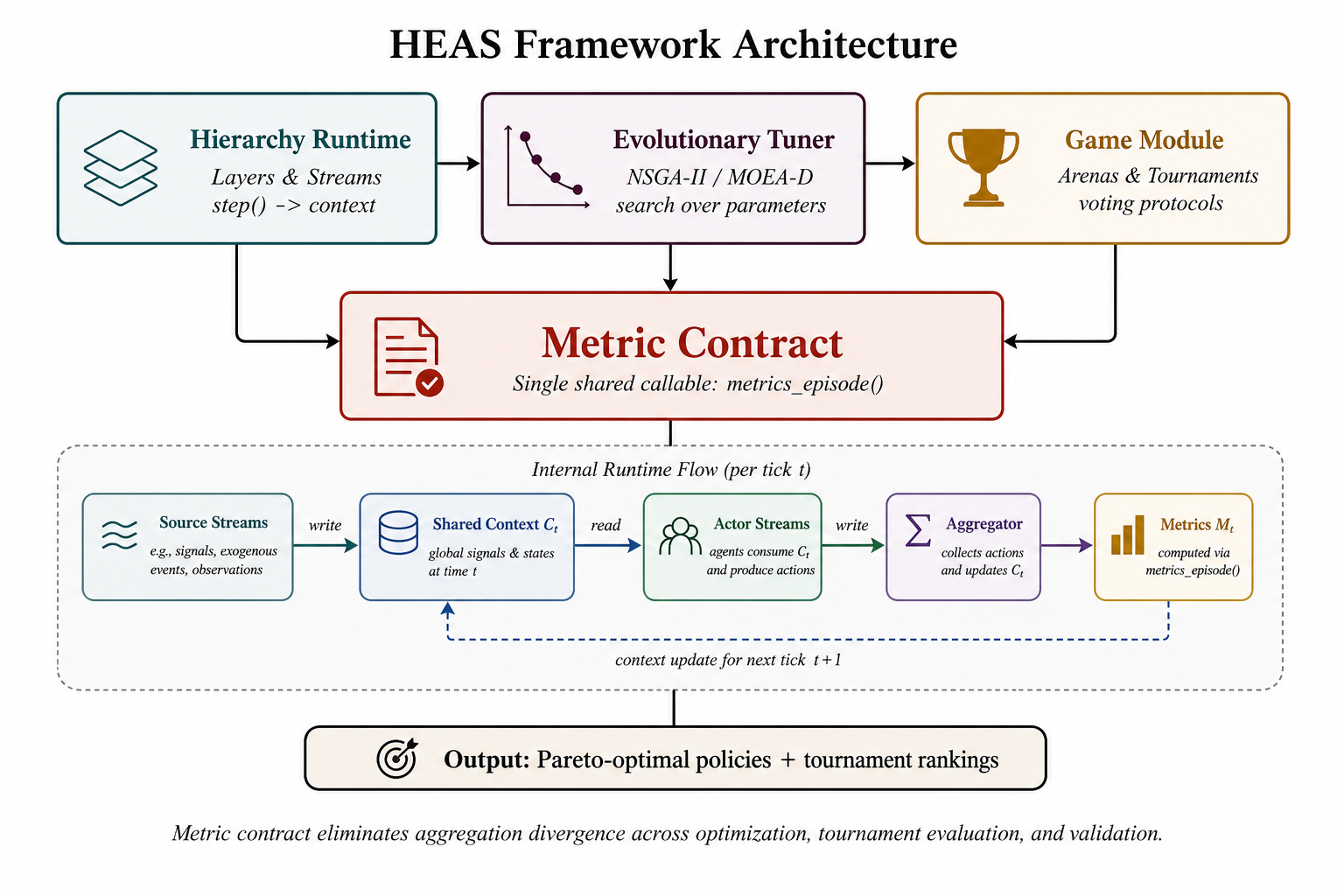}
\caption{HEAS three-module architecture. The Hierarchy Runtime composes simulations from layers and streams, the Evolutionary Tuner performs multi-objective search (NSGA-II/MOEA/D), and the Game Module evaluates policies across scenario ensembles. The Metric Contract ensures all three modules compute the same outcome metric through a single shared callable, eliminating aggregation divergence.}
\label{fig:architecture}
\end{figure}

\section{Use Cases}
HEAS has been applied to three reference studies that exercise different scientific logics while keeping the framework pipeline fixed. In each case, only the stream factories, gene schemas, scenarios, and episode metrics change; the hierarchy, search, tournament, and metric-contract interfaces remain the same. These studies demonstrate that the same framework surface supports both evolutionary search and held-out scenario evaluation without rewiring.

\textbf{Ecological population management} assembles a predator-prey arena with five streams, a 2-gene policy (\texttt{risk}, \texttt{dispersal}), and fragmentation $\times$ shock scenarios. The evolutionary tuner ran NSGA-II with a population of 200 over 100 generations (10 episodes per evaluation, seed 42), converging to a Pareto front of 6 non-dominated solutions with hypervolume 1.098. The game module then evaluated the Pareto-optimal policies across a held-out 64-scenario ensemble; the majority-vote champion (\texttt{risk}=0.00, \texttt{dispersal}=1.00) won all 64 scenarios, with zero rank reversal between optimization ranking and tournament ranking. The same model-construction and evaluation surface supported both stages without modification.

\textbf{Enterprise regulatory design} reuses the same framework contracts for a four-layer regulatory arena with a 4-gene policy (\texttt{tax\_rate}, \texttt{audit\_intensity}, \texttt{subsidy}, \texttt{penalty\_rate}). The tuner ran NSGA-II with a population of 300 over 150 generations (5 episodes per evaluation, seed 42), producing a large non-dominated set with hypervolume 4399.373. The champion policy (\texttt{tax\_rate}=0.50, \texttt{audit\_intensity}=0.17, \texttt{subsidy}=0.00, \texttt{penalty\_rate}=0.20) achieves welfare=34.49 with variance=84.90. The domain logic changes substantially, but the search and tournament wiring do not: the same interfaces carry a different objective pair, different streams, and a larger scenario ensemble.

\textbf{Wolf-Sheep ODE} keeps the framework interfaces fixed while swapping the underlying model family from an agent simulation to a mean-field Lotka-Volterra system. The tuner used a population of 100 over 80 generations (10 episodes per evaluation, seed 42), converging to a Pareto front of 2 solutions with hypervolume 5.498. This case is the clearest software portability test: the 4-layer arena required no additional framework-side coupling beyond a new \texttt{metrics\_episode()} implementation, confirming that the contract extends to a non-Mesa, non-stochastic system with the same HEAS pipeline.

\begin{figure}[htbp]
\centering
\includegraphics[width=0.98\linewidth]{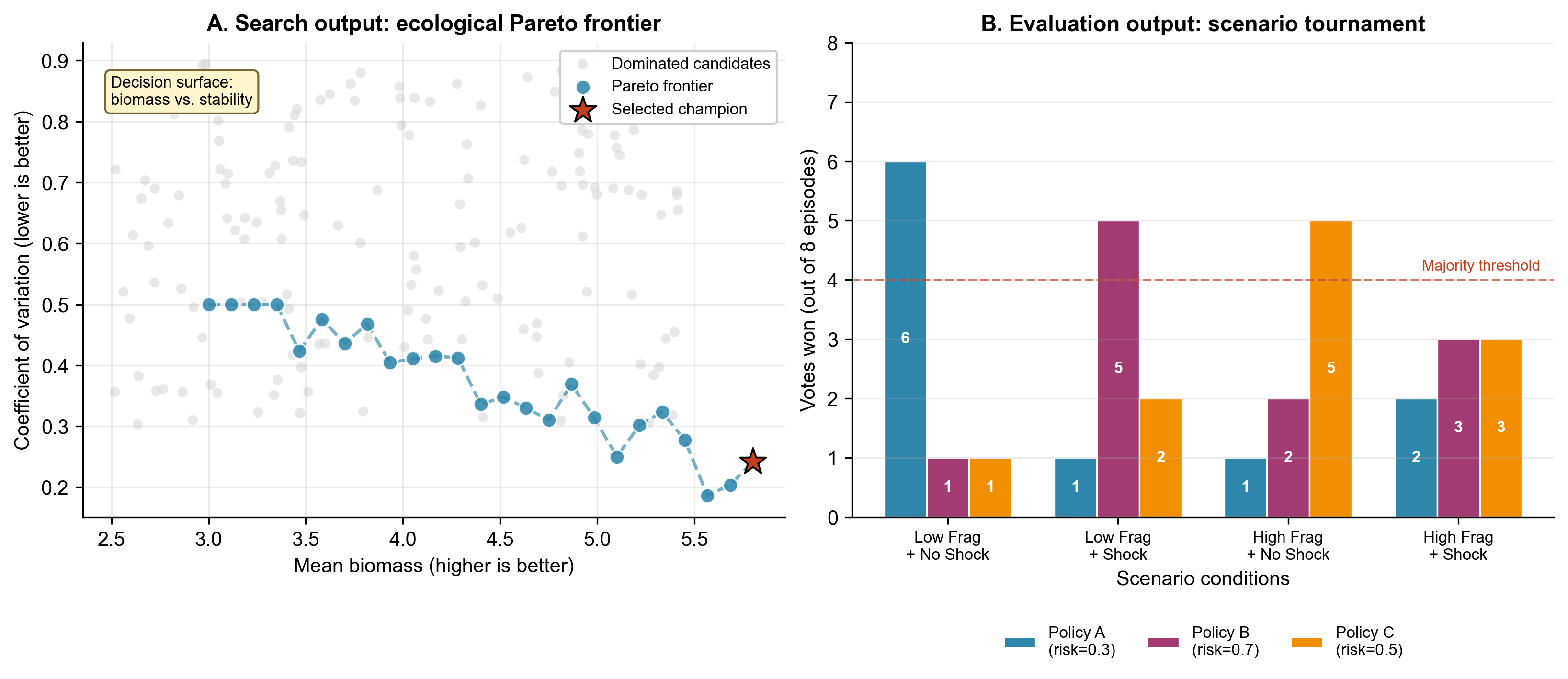}
\caption{Example outputs from the ecological arena case study. Left: Pareto front showing the trade-off between mean biomass (higher is better) and coefficient of variation (lower is better). The star marks the champion policy selected by HEAS. Right: Tournament voting summary showing how different policies perform across scenario conditions, with majority threshold indicated.}
\label{fig:pareto}
\end{figure}

\section{Availability and Reproducibility}
HEAS includes a command-line interface for batch runs and a browser-based playground at \url{https://ryzhanghason.github.io/heas/} for configuring simulations, inspecting Pareto fronts, and exporting publication bundles without a backend server. The repository includes a \texttt{CONTRIBUTING.md} guide, a \texttt{CODE\_OF\_CONDUCT.md}, release metadata, and automated tests. These materials are intended to make the software inspectable by reviewers and usable by researchers outside the original development team.

\section{Research Impact Statement}
HEAS has been developed as research software rather than as a one-off script. It is packaged for installation with \texttt{pip install heas}, released under the LGPL-3.0 license, documented with examples and a browser playground, and covered by 56 functional tests. Source code is available at \url{https://github.com/ryZhangHason/heas}.

\section{Acknowledgements}
The authors acknowledge the open-source simulation and optimization communities whose tools and documentation informed HEAS, especially Mesa, NetLogo, AgentPy, DEAP, EMA Workbench, and OpenMOLE.

\section{AI Usage Disclosure}
The authors used AI-assisted tools for code development, refactoring support and test scaffolding. Human authors reviewed, edited, and validated AI-assisted outputs, and the core problem framing, software architecture, research design, and final scholarly claims remain the responsibility of the authors.

\bibliography{paper}
\end{document}